\documentclass[runningheads,a4paper]{llncs}

\usepackage{amssymb,amsmath}
\usepackage{bm,bbm}
\usepackage{graphicx,subfigure}
\usepackage{booktabs}
\usepackage{microtype}

\usepackage{url}

\begin{document}

\title{Generalized Statistical Complexity of SAR Imagery}

\titlerunning{Generalized Statistical Complexity of SAR Imagery}

\author{Eliana S.\ de Almeida\inst{1}\and Antonio Carlos de Medeiros\inst{1}\and Osvaldo A.\ Rosso\inst{1,2}\and \mbox{Alejandro C.\ Frery\inst{1}}\thanks{The authors are grateful to CNPq and Fapeal for supporting this research}}

\authorrunning{Almeida, Medeiros, Rosso \& Frery}

\institute{
Universidade Federal de Alagoas -- UFAL\\
Laborat\'orio de Computa\c c\~ao Cient\'ifica e An\'alise Num\'erca -- LaCCAN\\
57072-970, Macei\'o, AL -- Brazil\and
Laboratorio de Sistemas Complejos\\
Facultad de Ingenier\'{\i}a\\
Universidad de Buenos Aires\\
Av.\ Paseo Col\'on 840, Ciudad Aut\'onoma de Buenos Aires, 1063 Argentina
}

\toctitle{Generalized Statistical Complexity of SAR Imagery}
\tocauthor{Almeida, Medeiros, Frery, Rosso}
\maketitle

\begin{abstract}
A new generalized Statistical Complexity Measure (SCM) was proposed by Rosso et al in 2010. It is a functional that captures the notions of order/disorder and of distance to an equilibrium distribution. The former is computed by a measure of entropy, while the latter depends on the definition of a stochastic divergence. When the scene is illuminated by coherent radiation, image data is corrupted by speckle noise, as is the case of ultrasound-B, sonar, laser and Synthetic Aperture Radar (SAR) sensors. In the amplitude and intensity formats, this noise is multiplicative and non-Gaussian requiring, thus, specialized techniques for image processing and understanding. One of the most successful family of models for describing these images is the Multiplicative Model which leads, among other probability distributions, to the $\mathcal G^0$ law. This distribution has been validated in the literature as an expressive and tractable model, deserving the ``universal'' denomination for its ability to describe most types of targets. In order to compute the statistical complexity of a site in an image corrupted by speckle noise, we assume that the equilibrium distribution is that of fully developed speckle, namely the Gamma law in intensity format, which appears in areas with little or no texture. We use the Shannon entropy along with the Hellinger distance to measure the statistical complexity of intensity SAR images, and we show that it is an expressive feature capable of identifying many types of targets.
\keywords{information theory, speckle, feature extraction}
\end{abstract}

\section{Introduction}
\label{sec:intro}

Synthetic Aperture Radar (SAR) is a prominent source of information for many Remote Sensing applications.
The data these devices provides carries information which is mostly absent in conventional sensors which operate in the optical spectrum or in its vicinity.
SAR sensors are active, in the sense that they carry their own illumination source and, therefore, are able to operate any time.
Since they operate in the microwaves region of the spectrum, they are mostly sensitive to the roughness and to the dielectric properties of the target.
The price to pay for these advantages is that these images are corrupted by a signal-dependent noise, called \textit{speckle}, which in the mostly used formats of SAR imagery is non-Gaussian and enters the signal in a non-additive manner.
This noise makes both automatic and visual analysis a hard task, and defies the use of classical features.

This paper presents a new feature for SAR image analysis called Generalized Statistical Complexity.
It was originally proposed and assessed for one-dimensional signals, for which it was shown to be able to detect transition points between different regimes~\cite{Rosso2007}.
This feature is the product of an entropy and a stochastic distance between the model which best describes the data and an equilibrium distribution~\cite{Lamberti2004,LopezRuiz1995}.

The statistical nature of speckled data allows to propose a Gamma law as the equilibrium distribution, while the $\mathcal G^0$ model describes the observed data with accuracy.
Both the entropy and the stochastic distance are derived within the framework of the so-called $(h,\phi)$ entropies and divergences, respectively, which stem from studies in Information Theory.

We show that the Statistical Complexity of SAR data, using the Shannon entropy and the Hellinger distance, stems as a powerful new feature for the analysis of this kind of data.

\section{The Multiplicative Model}
\label{sec:model}

The multiplicative model is one of the most successful frameworks for describing data corrupted by speckle noise.
It can be traced back to the work by Goodman~\cite{Goodman76}, where stems from the image formation being, therefore, phenomenological.
The multiplicative model for the intensity format states that the observation in every pixel is the outcome of a random variable $Z\colon\Omega\rightarrow\mathbb R_+$ which is the product of two independent random variables: $X\colon\Omega\rightarrow\mathbb R_+$, the ground truth or backscatter, related to the intrinsic dielectric properties of the target, and $Y\colon\Omega\rightarrow\mathbb R_+$, the speckle noise, obeying a unitary mean Gamma law.
The distribution of the return, $Z=XY$, is completely specified by the distributions $X$ and $Y$ obey.

The univariate multiplicative model began as a single distribution for the amplitude format, namely the Rayleigh law~\cite{JakemanPusey76}, was extended by Yueh et al.~\cite{yueh89} to accommodate the $K$ law and later improved further by Frery et al.~\cite{frery96} to the $G$ distribution, that generalizes all the previous probability distributions.
Gao~\cite{StatisticalModelingSARImagesSurvey} provides a complete and updated account of the distributions employed in the description of SAR data.

For the intensity format which we deal with in this article, the multiplicative model reduces to, essentially, two important distributions, namely the Gamma and the $\mathcal G^0$ laws.
The Gamma distribution is characterized by the density function
\begin{equation}
 f(z)=\frac{{(L/c)}^{L}}{\Gamma(L)}{z}^{L-1}\exp\{-Lz/c\},
\label{densgamacons}
\end{equation}
being $c>0$ the mean, $z>0$ and $L\geq1$, denoted $\Gamma(L,L/c)$. This is an adequate model for homogeneous regions as, for instance, pastures over flat relief.
The $\mathcal G^0$ law has density function
\begin{equation}
f(z)=\frac{L^{L}\Gamma(L-\alpha)}{\gamma^{\alpha}\Gamma(L)\Gamma(-\alpha)}\frac{z^{L-1}}{(\gamma+Lz)^{L-\alpha}},
\label{densga0}
\end{equation}
where $-\alpha,\gamma,z>0$, $L\geq1$, denoted $\mathcal G^{0}(\alpha,\gamma,L)$. This distribution was  proposed as a model for extremely heterogeneous areas~\cite{frery96}, and Mejail et al.~\cite{mejailfreryjacobobustos2001,MejailJacoboFreryBustos:IJRS} demonstrated it can be considered a universal model for speckled data.

Data obeying the $\Gamma$ law are referred to as ``fully developed speckle'', meaning that there is no texture in the wavelength of the illumination (which is in the order of centimeters).
The absolute value of the parameter $\alpha$ in Equation~(\ref{densgamacons}) is, on the contrary, a measure of the number of distinct objects of size of the order of the wavelength with which the scene is illuminated.
As $\alpha\to-\infty$, the $\mathcal G^0$ distribution becomes the $\Gamma$ law.

\section{Generalized Measure of Statistical Complexity}
\label{sec:measure}

The information content of a system  is typically evaluated via a probability distribution function (PDF) describing the apportionment of some measurable or observable quantity (i.e. a time series ${\mathcal S}(t)$). 
An information measure can primarily be viewed as a quantity that characterizes this given probability distribution $P$. 
The Shannon entropy is often used as a the ``natural" one~\cite{Shannon1949}. 
Given a discrete probability distribution $P = \{ p_i : i = 1, \cdots ,M \}$, with $M$ the degrees of freedom,  Shannon's logarithmic information measure reads ${\mathrm S}[P] = -\sum_{i=1}^{M}  p_i \ln( p_i)$.
It can be regarded as a measure of the uncertainty associated to the physical process described by $P$.
From now on we assume that the only restriction on the  PDF representing the state of our system is $\sum_{j= 1}^N p_j = 1$ (micro-canonical representation).
If ${\mathrm S}[P] = {\mathrm S}_{\min} = 0$ we are in position to predict with complete certainty which of the possible outcomes $i$, whose probabilities are given by $p_i$, will actually take place. 
Our knowledge of the underlying process described by the probability distribution is then maximal. 
In contrast, our knowledge is minimal for a uniform distribution and the uncertainty is maximal, ${\mathrm S}[P_e] = {\mathrm S}_{\max}$.

It is known that an entropic measure does not quantify the degree of structure or patterns present in a process~\cite{Feldman1998}. 
Moreover, it was recently shown that measures of statistical or structural complexity are necessary for a better understanding of chaotic time series  because they are able to capture their organizational properties~\cite{Feldman2008}. 
This kind of information is not revealed by measures of randomness. 
The extremes perfect order (like a periodic sequence) and maximal randomness (fair coin toss) possess no complex structure and exhibit zero statistical complexity.
There is a wide range of possible degrees of physical structure these extremes that should be quantified by
{\it statistical complexity measures.} 
Rosso and coworkers introduced an effective statistical complexity measure (SCM) that is able to detect essential details of the dynamics and differentiate different degrees of periodicity and chaos~\cite{Lamberti2004}. 
This specific SCM, abbreviated as MPR, provides important additional information regarding the peculiarities of the underlying probability distribution, not already detected by the entropy.

The statistical complexity measure is defined,  following the seminal, intuitive notion advanced by L\'opez-Ruiz  et al.
\cite{LopezRuiz1995}, via the product
\begin{equation}
C[P] =  H[P] \cdot D[P, P_{ref}].
\label{Complexity}
\end{equation}

The idea behind the Statistical Complexity is measuring at the same time the order/disorder of the system ($H$) and how far the system is from its equilibrium state (the so-called disequilibrium $D$)~\cite{GeneralizedStatisticalComplexityMeasuresGeometricalAnalyticalProperties,GeneralizedStatisticalComplexityMeasure}.
The first component can be obtained by means of an entropy, while the second requires computing a stochastic distance between the actual (observed) model and a reference one.
Salicr\'u et al.~\cite{salicruetal1993,Salicru1994} provide a very convenient conceptual framework for both of these measures.

Let $f_{\boldsymbol{Z}}(\boldsymbol{Z}';\boldsymbol{\theta})$ be a probability density function with parameter vector $\boldsymbol{\theta}$ which characterizes the distribution of the (possibly multivariate) random variable $\boldsymbol{Z}$.
The ($h,\phi$)-entropy relative to $\boldsymbol{Z}$ is defined by 
\begin{align*}
H_{\phi}^h(\boldsymbol{\theta})=h\Big(\int_{\mathcal A}\phi(f_{\boldsymbol{Z}}(\boldsymbol{Z}';\boldsymbol{\theta}))\mathrm{d}\boldsymbol{Z}'\Big),
\end{align*}
where either $\phi:\bigl[0,\infty\bigr) \rightarrow \mathbb{R}$ is concave and $h:\mathbb{R} \rightarrow \mathbb{R}$ is increasing, or $\phi$ is convex and $h$ is decreasing.   
The differential element $\mathrm{d}\boldsymbol{Z}'$ sweeps the whole support $\mathcal A$.
In this work we only employ the Shannon entropy, for which $h(y)=y$ and $\phi(x)=-x\ln x$.

Consider now the (possibly multivariate) random variables $\boldsymbol{X}$ and $\boldsymbol{Y}$ with densities $f_{\boldsymbol{X}}(Z;\boldsymbol{\theta_1})$ and $f_{\boldsymbol{Y}}(Z;\boldsymbol{\theta_2})$, respectively, where $\boldsymbol{\theta_1}$ and $\boldsymbol{\theta_2}$ are parameter vectors.
The densities are assumed to have the same support $\boldsymbol{\mathcal A}$.
The $(h,\phi)$-divergence between $f_{\boldsymbol{X}}$ and $f_{\boldsymbol{Y}}$ is defined by
\begin{equation} 
D_{\phi}^h(\boldsymbol{X},\boldsymbol{Y}) = 
h\biggl(\int_{\boldsymbol{\mathcal A}} \phi\biggl( \frac{f_{\boldsymbol{X}}({Z};\boldsymbol{\theta_1})}{f_{\boldsymbol{Y}}({Z};\boldsymbol{\theta_2})}\biggr) f_{\boldsymbol{Y}}({Z};\boldsymbol{\theta_2})\mathrm{d}{Z}\biggr),
\label{eq:eps2-no}
\end{equation}
where $h\colon(0,\infty)\rightarrow[0,\infty)$ is a strictly increasing function with $h(0)=0$ and $\phi\colon (0,\infty)\rightarrow[0,\infty)$ is a convex function such that $0\,\phi(0/0)=0$ and $0\,\phi(x/0)=\lim_{x \rightarrow \,\infty} \phi(x)/x$. 
The differential element $\mathrm{d}{Z}$ sweeps the support.
In the following we will only employ the Hellinger divergence which is also a distance, for which $h(y)={y}/{2}$, $0\leq y<2$ and $\phi(x)=(\sqrt{x}-1)^2$.

The influence of the choice of a distance when computing statistical complexities is studied in Reference~\cite{DistancesProbabilitySpaceStatisticalComplexitySetup}.
Following Rosso et al.~\cite{GeneralizedStatisticalComplexityMeasure}, we work with the Hellinger distance and we define the Statistical Complexity of coordinate $(i,j)$ in an intensity SAR image as the product
\begin{equation}
C(i,j) = H(i,j) \cdot D(i,j), \label{eq:StatisticalComplexity}
\end{equation}
where $H(i,j)$ is the Shannon entropy observed in $(i,j)$ under the $\mathcal G^0$ model, and $D(i,j)$ is the observed Hellinger distance between the universal model (the $\mathcal G^0$ distribution) and the reference model of fully developed speckle (the $\Gamma$ law).

As previously noted, if an homogeneous area is being analyzed, the $\mathcal{G}^0$ and $\Gamma$ model can be arbitrarily close, and the distance between them tends to zero.
The entropy of the $\mathcal{G}^0$ model is closely related to the roughness of the target, as will be seen later, that is measured by $\alpha$.

Computing these observed quantities requires the estimation of the parameters which characterize the $\Gamma$ distribution ($c$, the sample mean) and the $\mathcal G^0$ law ($\alpha$ and $\gamma$), provided the number of looks $L$ is known.
The former is immediate, while estimating the later by maximum likelihood requires solving a nonlinear optimization problem.
The estimation is done using data in a vicinity of $(i,j)$.
Once obtained $\widehat{c}$ and $(\widehat{\alpha},\widehat{\gamma})$, the terms in Equation~\eqref{eq:StatisticalComplexity} are computed by numerical integration.
References~\cite{AllendeFreryetal:JSCS:05,FreryCribariSouza:JASP:04} discuss venues for estimating the parameters of the $\mathcal{G}^0$ law safely.


\section{Results}\label{sec:assessment}

\begin{figure}[hbt]
\centering
\subfigure[Original E-SAR image\label{fig:OriginalImage}]{\includegraphics[width=.49\linewidth]{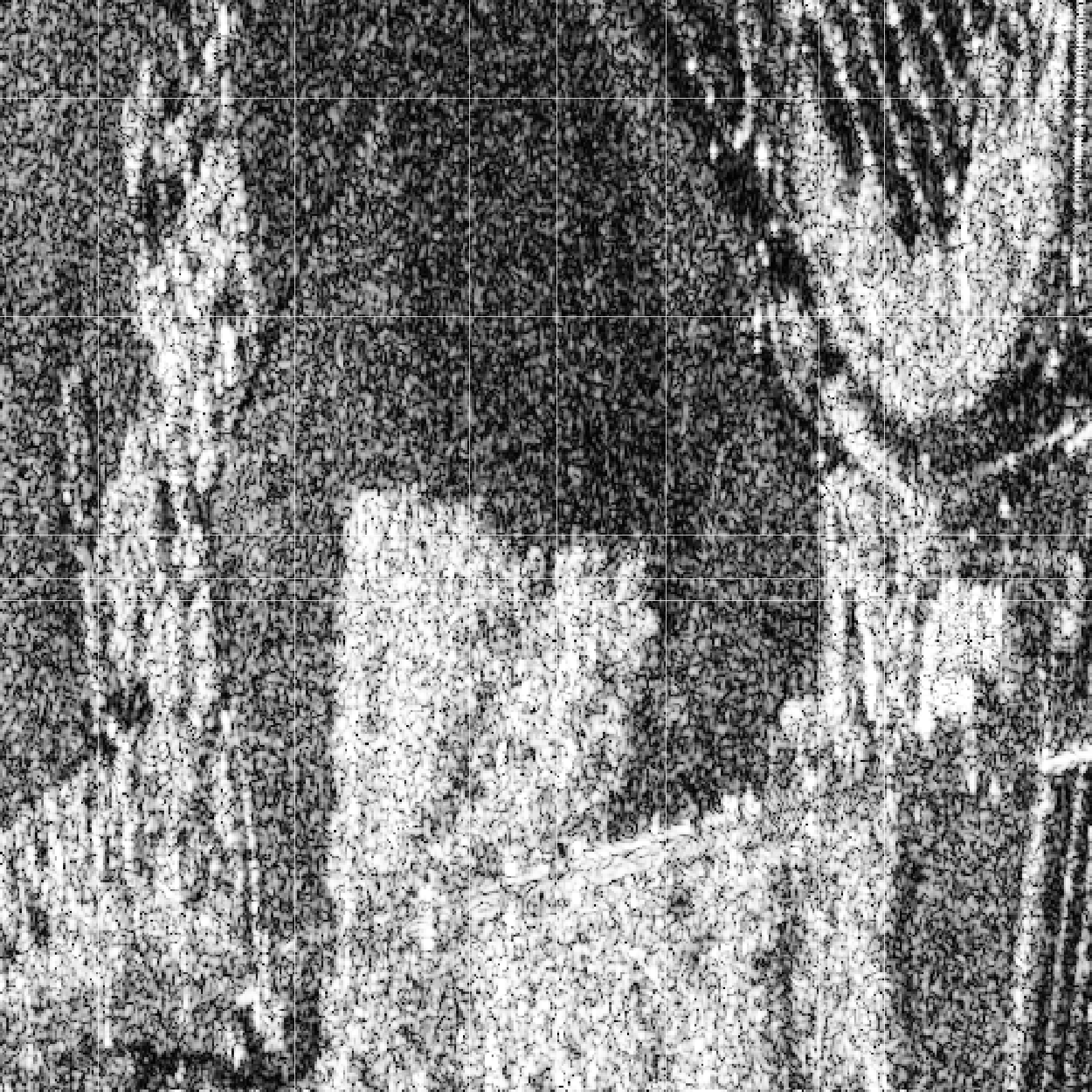}}
\subfigure[Entropies\label{fig:Entropies}]{\includegraphics[width=.49\linewidth]{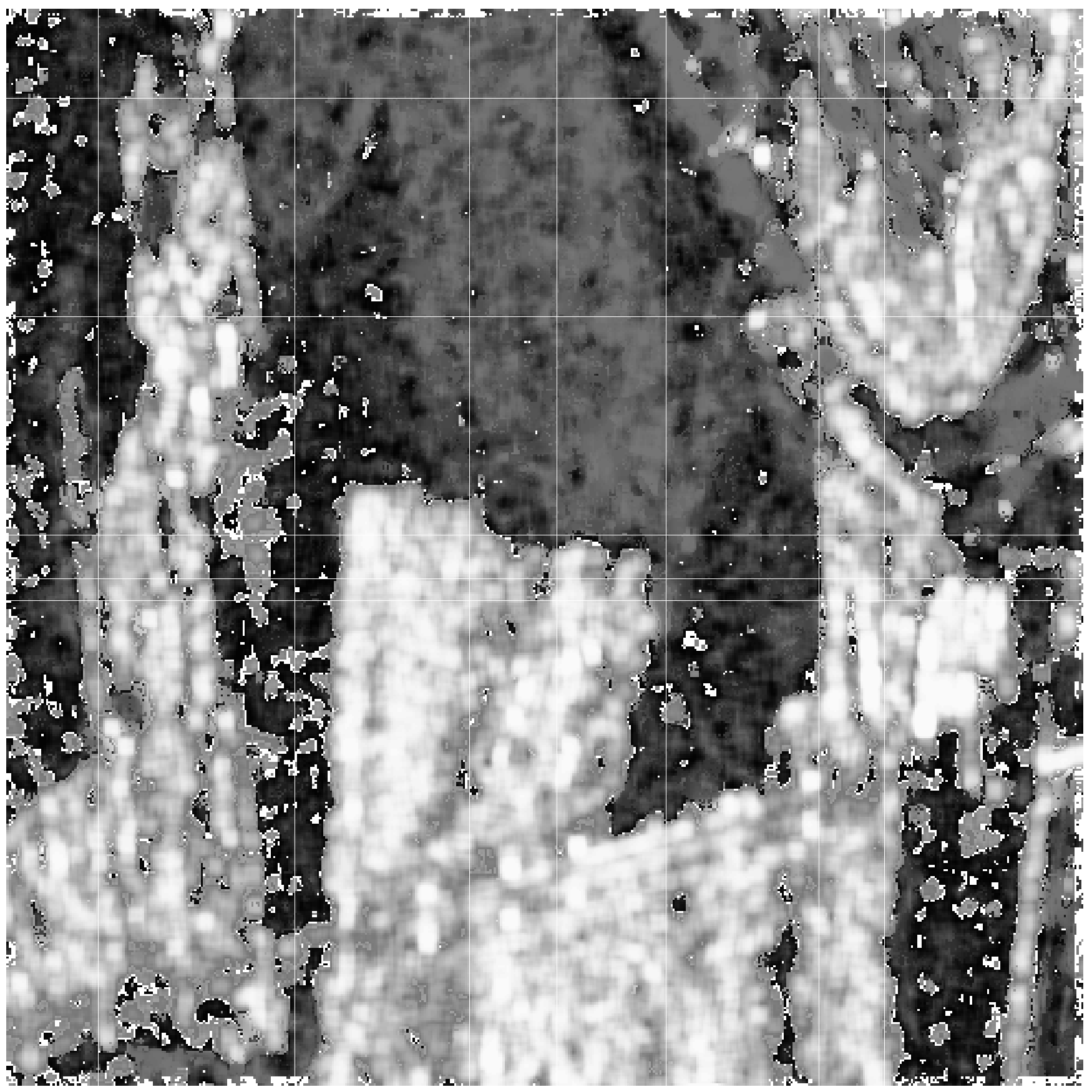}}\\
\subfigure[Hellinger distances\label{fig:Distances}]{\includegraphics[width=.49\linewidth]{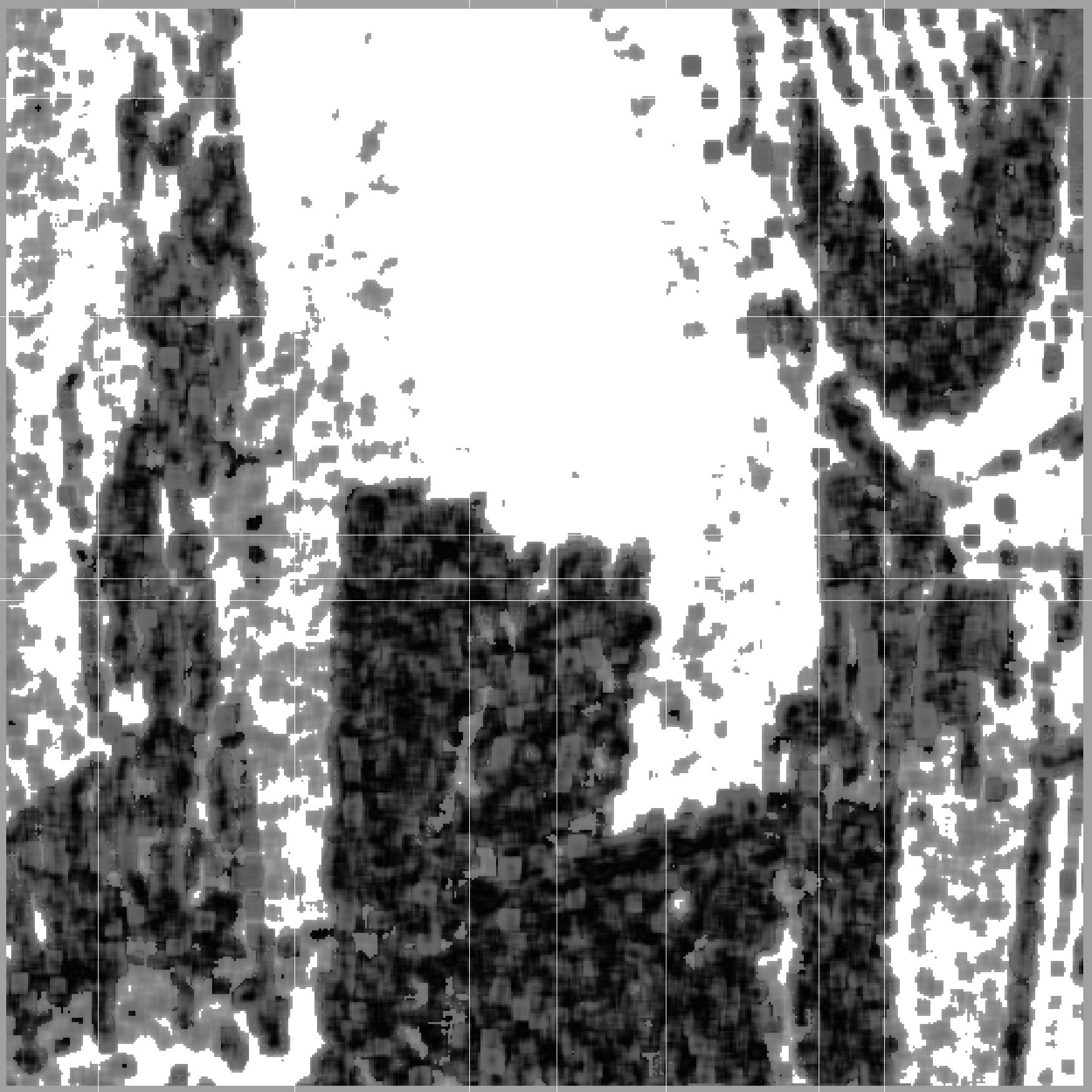}}
\subfigure[Statistical Complexities\label{fig:Complexities}]{\includegraphics[width=.49\linewidth]{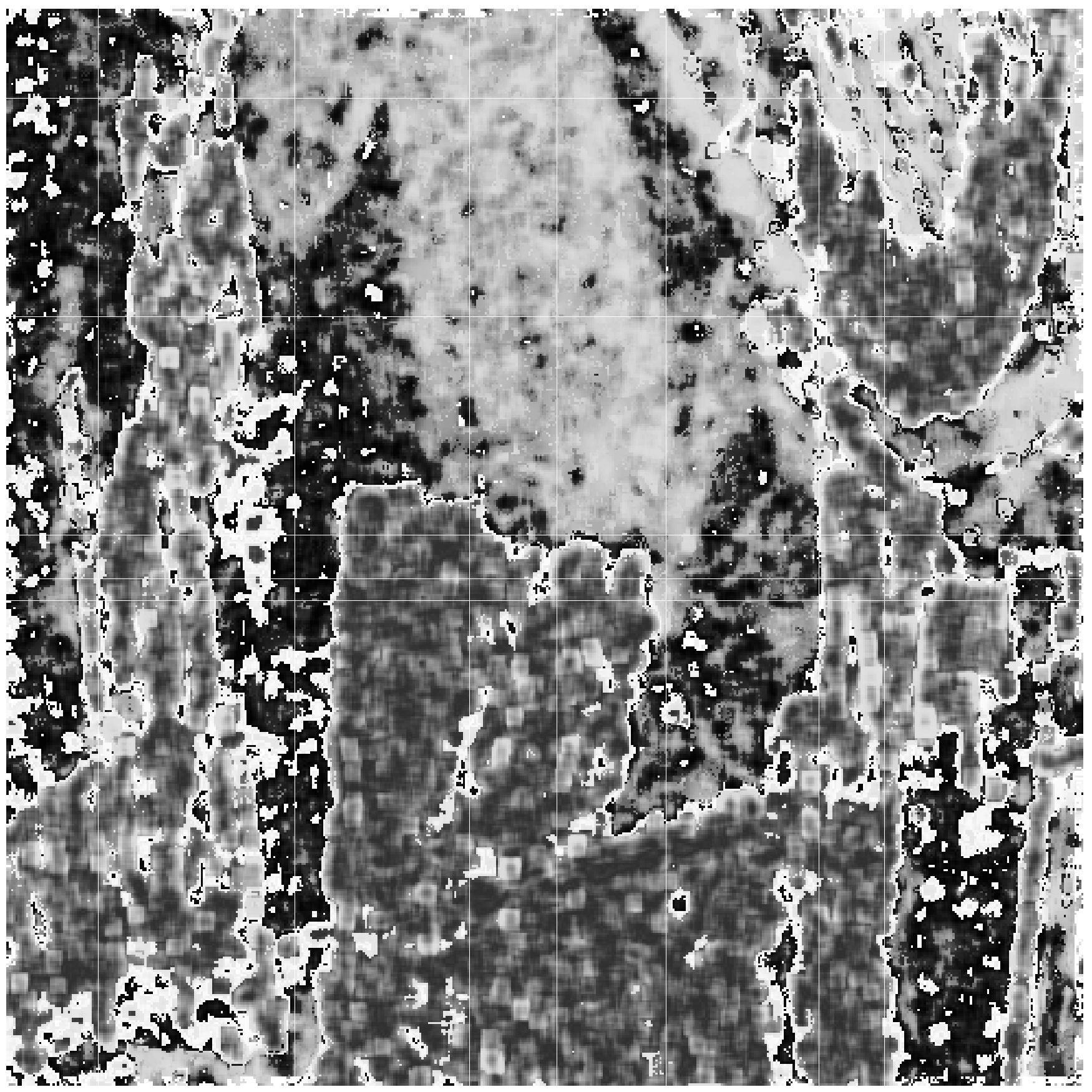}}
\caption{Results of applying the proposed feature extraction to an E-SAR image over Munich, Germany}\label{fig:Results}
\end{figure}

Figure~\ref{fig:Results} presents the main results obtained with the proposed measures.
Figure~\ref{fig:OriginalImage} shows the original image which was obtained by the E-SAR sensor, an airborne experimental polarimetric SAR, over Munich, Germany.
Only the intensity HH channel is employed in this study.
The image was acquired with three nominal looks.
The scene consists mostly of different types of crops (the dark areas), forest and urban areas (the bright targets).
\ref{densga0}
Figure~\ref{fig:Entropies} shows the Shannon entropy as shades of gray whose brightness is proportional to the observed value.
It is remarkable that this measure is closely related to the roughness of the target, i.e., the brighter the pixel the more heterogeneous the area.
The entropy is also able to discriminate between different types of homogeneous targets, as shown in the various types of dark shades.

Figure~\ref{fig:Distances} shows the Hellinger distance between the universal model and the model for fully developed speckle.
As expected, the darkest values are related to areas of low level of roughness, while the brightest spots are the linear strips in the uppermost right corner, since they are man-made structures.

The Statistical Complexity is shown in Figure~\ref{fig:Complexities}.
It summarizes the evidence provided by the entropy (Figure~\ref{fig:Entropies}) and by the stochastic distance between models (Figure~\ref{fig:Distances}).
As it can be seen in the image, the values exhibit more variation than their constituents, allowing a fine discrimination of targets.
As such, it stems as a new and relevant feature for SAR image analysis.

The data were read, processed, analyzed and visualized using \texttt{R} v.~2.14.0 \cite{R} on a MacBook Pro running Mac OS X v.~10.7.3.
This platform is freely available at \url{http://www.r-project.org} for a diversity of computational platforms, and its excellent numerical properties have been attested in \cite{OctaveScilabMatlabCAM,AlmironSilvaMM:2009}.

\section{Conclusions}\label{sec:conclu}

The Statistical Complexity of SAR images reveals information which is not available either through the mean (which is the parameter of the model for homogeneous areas) or by the parameters of the model for extremely heterogeneous areas.
As such, it appears as a promising feature for SAR image analysis.

Ongoing studies include the derivation of 	analytical expressions for the entropy and the Hellinger distance, other stochastic distances, the sample properties of the Statistical Complexity and its generalization for other models including Polarimetric SAR.

\end{document}